\documentclass[a4paper,11pt]{article}
\usepackage[T1]{fontenc} % if needed
\usepackage{subfigure}
\makeatletter
\gdef\@fpheader{ }
\gdef\@journal{ }
\makeatother
\RequirePackage{amsmath}
\RequirePackage{amssymb}
\RequirePackage{epsfig}
\RequirePackage{graphicx}
\RequirePackage[numbers,sort&compress]{natbib}
\RequirePackage{color}
\RequirePackage[colorlinks=true
,urlcolor=blue
,anchorcolor=blue
,citecolor=blue
,filecolor=blue
,linkcolor=blue
,menucolor=blue
,pagecolor=blue
,linktocpage=true
,pdfproducer=medialab
,pdfa=true
]{hyperref}

\newif\ifnotoc\notocfalse
\newif\ifemailadd\emailaddfalse
\newif\iftoccontinuous\toccontinuousfalse
\makeatletter
\def\@subheader{\@empty}
\def\@keywords{\@empty}
\def\@abstract{\@empty}
\def\@xtum{\@empty}
\def\@dedicated{\@empty}
\def\@arxivnumber{\@empty}
\def\@collaboration{\@empty}
\def\@collaborationImg{\@empty}
\def\@proceeding{\@empty}
\def\@preprint{\@empty}

\newcommand{\subheader}[1]{\gdef\@subheader{#1}}
\newcommand{\keywords}[1]{\if!\@keywords!\gdef\@keywords{#1}\else%
\PackageWarningNoLine{\jname}{Keywords already defined.\MessageBreak Ignoring last definition.}\fi}
\renewcommand{\abstract}[1]{\gdef\@abstract{#1}}
\newcommand{\dedicated}[1]{\gdef\@dedicated{#1}}
\newcommand{\arxivnumber}[1]{\gdef\@arxivnumber{#1}}
\newcommand{\proceeding}[1]{\gdef\@proceeding{#1}}
\newcommand{\xtumfont}[1]{\textsc{#1}}
\newcommand{\correctionref}[3]{\gdef\@xtum{\xtumfont{#1} \href{#2}{#3}}}
\newcommand\jname{JHEP}
\newcommand\acknowledgments{\section*{Acknowledgments}}

\newcommand\preprint[1]{\gdef\@preprint{\hfill #1}}

\makeatother

\newcommand\note[2][]{%
\if!#1!%
\stepcounter{footnote}\footnotetext{#2}%
\else%
{\renewcommand\thefootnote{#1}%
\footnotetext{#2}}%
\fi}

\makeatletter
\newtoks\auth@toks
\renewcommand{\author}[2][]{%
  \if!#1!%
    \auth@toks=\expandafter{\the\auth@toks#2\ }%
  \else
    \auth@toks=\expandafter{\the\auth@toks#2$^{#1}$\ }%
  \fi
}
\makeatother
\makeatletter
\newtoks\affil@toks\newif\ifaffil\affilfalse
\newcommand{\affiliation}[2][]{%
\affiltrue
  \if!#1!%
    \affil@toks=\expandafter{\the\affil@toks{\item[]#2}}%
  \else
    \affil@toks=\expandafter{\the\affil@toks{\item[$^{#1}$]#2}}%
  \fi
}
\makeatother
\makeatletter
\newtoks\email@toks\newcounter{email@counter}%
\newcommand{\emailAdd}[1]{%
\emailaddtrue%
\ifnum\theemail@counter>0\email@toks=\expandafter{\the\email@toks, \@email{#1}}%
\else\email@toks=\expandafter{\the\email@toks\@email{#1}}%
\fi\stepcounter{email@counter}}
\newcommand{\@email}[1]{\href{mailto:#1}{\tt #1}}
\makeatother

\makeatletter
\newcommand*\collaboration[1]{\gdef\@collaboration{#1}}
\newcommand*\collaborationImg[2][]{\gdef\@collaborationImg{#2}}
\makeatletter
\newcommand\afterLogoSpace{\smallskip}
\newcommand\afterSubheaderSpace{\vskip3pt plus 2pt minus 1pt}
\newcommand\afterProceedingsSpace{\vskip21pt plus0.4fil minus15pt}
\newcommand\afterTitleSpace{\vskip23pt plus0.06fil minus13pt}
\newcommand\afterRuleSpace{\vskip23pt plus0.06fil minus13pt}
\newcommand\afterCollaborationSpace{\vskip3pt plus 2pt minus 1pt}
\newcommand\afterCollaborationImgSpace{\vskip3pt plus 2pt minus 1pt}
\newcommand\afterAuthorSpace{\vskip5pt plus4pt minus4pt}
\newcommand\afterAffiliationSpace{\vskip3pt plus3pt}
\newcommand\afterEmailSpace{\vskip16pt plus9pt minus10pt\filbreak}
\newcommand\afterXtumSpace{\par\bigskip}
\newcommand\afterAbstractSpace{\vskip16pt plus9pt minus13pt}
\newcommand\afterKeywordsSpace{\vskip16pt plus9pt minus13pt}
\newcommand\afterArxivSpace{\vskip3pt plus0.01fil minus10pt}
\newcommand\afterDedicatedSpace{\vskip0pt plus0.01fil}
\newcommand\afterTocSpace{\bigskip\medskip}
\newcommand\afterTocRuleSpace{\bigskip\bigskip}
\newlength{\affiliationsSep}
\newcommand\beforetochook{\pagestyle{myplain}\pagenumbering{roman}}

\DeclareFixedFont\trfont{OT1}{phv}{b}{sc}{11}

\renewcommand\maketitle{
\pagestyle{empty}
\thispagestyle{titlepage}
\setcounter{page}{0}
\noindent{\small\scshape\@fpheader}\@preprint\par

\afterLogoSpace
\if!\@subheader!\else\noindent{\trfont{\@subheader}}\fi
\afterSubheaderSpace
\if!\@proceeding!\else\noindent{\sc\@proceeding}\fi
\afterProceedingsSpace
{\LARGE\flushleft\sffamily\bfseries\@title\par}
\afterTitleSpace
\hrule height 1.5\p@%
\afterRuleSpace
\if!\@collaboration!\else
{\Large\bfseries\sffamily\raggedright\@collaboration}\par
\afterCollaborationSpace
\fi
\if!\@collaborationImg!\else
{\normalsize\bfseries\sffamily\raggedright\@collaborationImg}\par
\afterCollaborationImgSpace
\fi
{\bfseries\raggedright\sffamily\the\auth@toks\par}
\afterAuthorSpace
\ifaffil\begin{list}{}{%
\setlength{\leftmargin}{0.28cm}%
\setlength{\labelsep}{0pt}%
\setlength{\itemsep}{\affiliationsSep}%
\setlength{\topsep}{-\parskip}}
\itshape\small%
\the\affil@toks
\end{list}\fi
\afterAffiliationSpace
\ifemailadd %% if emailadd is true
\noindent\hspace{0.28cm}\begin{minipage}[l]{.9\textwidth}
\begin{flushleft}
\textit{E-mail:} \the\email@toks
\end{flushleft}
\end{minipage}
\else %% if emailaddfalse do nothing
\PackageWarningNoLine{\jname}{E-mails are missing.\MessageBreak Plese use \protect\emailAdd\space macro to provide e-mails.}
\fi
\afterEmailSpace
\if!\@xtum!\else\noindent{\@xtum}\afterXtumSpace\fi
\if!\@abstract!\else\noindent{\renewcommand\baselinestretch{.9}\textsc{Abstract:}}\ \@abstract\afterAbstractSpace\fi
\if!\@keywords!\else\noindent{\textsc{Keywords:}} \@keywords\afterKeywordsSpace\fi
\if!\@arxivnumber!\else\noindent{\textsc{ArXiv ePrint:}} \href{http://arxiv.org/abs/\@arxivnumber}{\@arxivnumber}\afterArxivSpace\fi
\if!\@dedicated!\else\vbox{\small\it\raggedleft\@dedicated}\afterDedicatedSpace\fi
\ifnotoc\else
\iftoccontinuous\else\newpage\fi
\beforetochook\hrule
\tableofcontents
\afterTocSpace
\hrule
\afterTocRuleSpace
\fi
\setcounter{footnote}{0}
\pagestyle{myplain}\pagenumbering{arabic}
} % close the \renewcommand\maketitle{

\renewcommand{\baselinestretch}{1.1}\normalsize
\divide\@tempdima\baselineskip
\@tempcnta=\@tempdima
\skip\@mpfootins = \skip\footins
\renewcommand{\@dotsep}{10000}

\newcommand\ps@myplain{
\pagenumbering{arabic}
\renewcommand\@oddfoot{\hfill-- \thepage\ --\hfill}
\renewcommand\@oddhead{}}
\let\ps@plain=\ps@myplain

\newcommand\ps@titlepage{\renewcommand\@oddfoot{}\renewcommand\@oddhead{}}

\numberwithin{equation}{section}

\renewcommand\section{\@startsection{section}{1}{\z@}%
                                   {-3.5ex \@plus -1.3ex \@minus -.7ex}%
                                   {2.3ex \@plus.4ex \@minus .4ex}%
                                   {\normalfont\large\bfseries}}
\renewcommand\subsection{\@startsection{subsection}{2}{\z@}%
                                   {-2.3ex\@plus -1ex \@minus -.5ex}%
                                   {1.2ex \@plus .3ex \@minus .3ex}%
                                   {\normalfont\normalsize\bfseries}}
\renewcommand\subsubsection{\@startsection{subsubsection}{3}{\z@}%
                                   {-2.3ex\@plus -1ex \@minus -.5ex}%
                                   {1ex \@plus .2ex \@minus .2ex}%
                                   {\normalfont\normalsize\bfseries}}
\renewcommand\paragraph{\@startsection{paragraph}{4}{\z@}%
                                   {1.75ex \@plus1ex \@minus.2ex}%
                                   {-1em}%
                                   {\normalfont\normalsize\bfseries}}
\renewcommand\subparagraph{\@startsection{subparagraph}{5}{\parindent}%
                                   {1.75ex \@plus1ex \@minus .2ex}%
                                   {-1em}%
                                   {\normalfont\normalsize\bfseries}}

\def\fnum@figure{\textbf{\figurename\nobreakspace\thefigure}}
\def\fnum@table{\textbf{\tablename\nobreakspace\thetable}}

\long\def\@makecaption#1#2{%
  \vskip\abovecaptionskip
  \sbox\@tempboxa{\small #1. #2}%
  \ifdim \wd\@tempboxa >\hsize
    \small #1. #2\par
  \else
    \global \@minipagefalse
    \hb@xt@\hsize{\hfil\box\@tempboxa\hfil}%
  \fi
  \vskip\belowcaptionskip}

\renewenvironment{thebibliography}[1]{%
\begin{oldthebibliography}{#1}%
\small%
\raggedright%
\setlength{\itemsep}{5pt plus 0.2ex minus 0.05ex}%
}%
{%
\end{oldthebibliography}%
}

\begin{document}

%%%%%%%%%%%%%%%%%%%%%%%%%%%%%%%%%%%%%%%%%%%%%%%%%%%%%%%%%%%%%%%%%%%%%%%%%%%%%%%%%%

\title{\boldmath Statistical problem of ideal gas in general 2-dimensional
regions}

%% %simple case: 2 authors, same institution
%% \author{A. Uthor}
%% \author{and A. Nother Author}
%% \affiliation{Institution,\\Address, Country}

% more complex case: 4 authors, 3 institutions, 2 footnotes
\author[a,1]{Ci Song, }\note{cisong@mail.nankai.edu.cn}
\author[b,2]{Wen-Du Li, }\note{lwd@tju.edu.cn}
\author[b]{Pardon Mwansa, }
\author[c,d,3]{and Ping Zhang}\note{zhangping@cueb.edu.cn}

% The "\note" macro will give a warning: "Ignoring empty anchor..."
% you can safely ignore it.

\affiliation[a]{Theoretical Physics Division, Chern Institute of Mathematics, Nankai University, Tianjin, 300071, P.R.China}
\affiliation[b]{Department of Physics, Tianjin University, Tianjin 300072, P.R.China}
\affiliation[c]{School of Finance, Capital University of Economics and Business, Beijing 100070, P.R.China}\emph{}
\affiliation[d]{Institute of Quantitative Finance and Risk Management, Capital University of Economics and Business, Beijing 100070, P.R.China}
% e-mail addresses: one for each author, in the same order as the authors
%\emailAdd{songci@ntu.cn}
%\emailAdd{second@asas.edu}
%\emailAdd{third@one.univ}
%\emailAdd{fourth@one.univ}

\abstract{In this paper, based on the conformal mapping method and the perturbation theory, we develop a
method to solve the statistical problem within general 2-dimensional regions. We consider some examples and the
numerical results and fitting results are given. We also give the thermodynamic quantities of the general 2-dimensional
regions, and compare the thermodynamic quantities of the different regions. }


\maketitle
\flushbottom
%%%%%%%%%%%%%%%%%%%%%%%%%%%%%%%%%%%%%%%%%%%%%%%%%%%%%%%%%%%%%%%%
\section{Introduction}

In this paper, we develop an approach to solve the statistical problem of an ideal gas in
general 2-dimensional regions based on the conformal mapping method. Using the conformal
mapping method and perturbation theory in Ref. \cite{robnik1984quantising}, we calculate
the spectrum of the ideal gas, and then we provide the partition function of the ideal gas
on a general 2-dimensional region.

It is difficult to solve the statistical problem within a general
2-dimensional region. The calculation of the partition function requires the
spectrum of the ideal gas. In order to calculate the spectrum of the ideal gas
in a general 2-dimensional region, we should consider a quantum
billiard problem, which is defined as a region of space where a point particle is
able to move freely, not being allowed to escape. We should calculate the eigenvalue
of Schr\"{o}dinger equation in this case. For a general 2-dimensional region,
the calculation is very difficult. In fact, the exact solutions only exist for some
special case, such as rectangular region, circular region
\cite{amore2010spectroscopy}, etc. We expect to develop an approach that can
solve the statistical problem of the ideal gas with in a general 2-dimensional region
based on some of these exact solutions.

There are some different approaches that have been developed to solve the
Schr\"{o}dinger equation within general 2-dimensional regions, you can get a
review of this problem in the Ref. \cite{amore2010spectroscopy}. Among these
approaches, the conformal mapping method is the most pronounced method, which is map
the original region onto a given region by a conformal
transformation \cite{robnik1984quantising}. The boundary condition can be
simplified by the conformal transformation, but the equation of motion become
complicated. Further calculation of the eigenvalues requires the use of perturbation
theory. So based on the approach elaborated in Ref. \cite{segel1961application, segel1961application2, robnik1984quantising},
we can calculate the eigenvalue on the general 2-dimensional region. Using the spectrum of the
ideal gas, we can obtain the partition function of the ideal gas.

The conformal mapping method is very important in mathematical physics. We
know, that if the transformation is a holomorphic function defined on a region of
the complex plane, then the transformation is a conformal map, which
maps each point belonging to the region specified onto another region. With the
conformal mapping method, we can often simplify some physical problems.

Now, we briefly review the conformal mapping method. Let us consider a Laplace
equation over a 2-dimensional region $D$,%
\begin{equation}
\phi_{xx}+\phi_{yy}=0,
\end{equation}
with Dirichlet boundary conditions. The points of the region $D$ are denoted
by $z\left(  x,y\right)  $. According to the Riemann mapping
theorem \cite{kodaira2007complex}, we can map the region $D$ onto a given
region $\Omega$ with simple boundary by a conformal transformation
$\omega=f\left(  z\right)  $, i.e.
\begin{equation}
\omega=u\left(  x,y\right)  +iv\left(  x,y\right)  =f\left(  z\right)  ,
\end{equation}
where $z=x+iy$. Under the conformal transformation, the equation transformed
as
\begin{equation}
\left(  u_{x}^{2}+u_{y}^{2}\right)  \phi_{uu}+2\left(  u_{x}v_{x}+u_{y}%
v_{y}\right)  \phi_{uv}+\left(  v_{x}^{2}+v_{y}^{2}\right)  \phi_{vv}+\left(
u_{xx}+u_{yy}\right)  \phi_{u}+\left(  v_{xx}+v_{yy}\right)  \phi_{v}=0.
\end{equation}
Because the transformation $\omega=f\left(  z\right)  $ is a holomorphic
function, $u\left(  x,y\right)  $ and $v\left(  x,y\right)  $ satisfy the
Cauchy--Riemann condition, i.e.%
\begin{equation}
\begin{cases}
    \frac{\partial u}{\partial x}= \frac{\partial v}{\partial y},\\
    \frac{\partial u}{\partial y}= -\frac{\partial v}{\partial x}.
\end{cases}
\end{equation}
With Cauchy--Riemann condition, the equation on the new region can be
simplified,
\begin{equation}
\left\vert f^{\prime}\left(  z\right)  \right\vert ^{2}\left(  \phi_{uu}%
+\phi_{vv}\right)  =0.
\end{equation}
Obviously, this is still a Laplace equation, except for a conformal factor
$\left\vert f^{\prime}\left(  z\right)  \right\vert ^{2}$.

If we consider a Poisson equation%
\begin{equation}
\phi_{xx}+\phi_{yy}=f\left(  x,y\right)  ,
\end{equation}
we can deal with this equation in the same manor as we did the Laplace equation. After a conformal transformation
$\omega=f\left(  z\right)  $, we obtain the equation on the new region, i.e.
\begin{equation}
\phi_{uu}+\phi_{vv}=\frac{1}{\left\vert f^{\prime}\left(  z\right)
\right\vert ^{2}}f\left[  x\left(  u,v\right)  ,y\left(  u,v\right)  \right]
.
\end{equation}

In Ref. \cite{segel1961application,segel1961application2,robnik1984quantising,amore2010spectroscopy}, perturbation
theory based on the conformal mapping method has been applied to solve the
eigenvalue problem on general 2-dimensional regions. In this paper, we also
calculate the partition function of ideal gas on general 2-dimensional regions
by adopting the same approach. We will give numerical results of the partition function.
Moreover the results of data fitting will be provided.

The statistical thermodynamics of the ideal gas confined in a finite domain
has been widely studied. In Ref. \cite{sisman2004casimir, sisman2004surface},
the authors calculate the size effect of ideal gases confined in a finite
domain due to the wave character of atoms. For an ideal gas confined in a
finite domain, there is a layer in which the density goes to zero which arises as a consequence
arises as a consequence of the size effect.
The boundary layer has been discussed in the Ref.
\cite{sisman2007quantum, firat2009universality}. In Ref. \cite{dai2003quantum,
dai2004geometry}, the authors considered the effects of boundary and
connectivity on ideal quantum gases in a confined space. In Ref.
\cite{dai2005hard, dai2007interacting}, the authors discuss the properties of
the hard-sphere gases in finite-size containers and convert the effect of
interparticle hard-sphere interaction to a kind of boundary effect. In Ref.
\cite{pang2006difference}, the authors analyze the difference between
boundary effects on Bosonic and Fermionic systems. Due to the size effect and
boundary effect, we have to consider the non-extensive effect. The general
discussion on the non-extensive effect can be found in \cite{huang2009inherent}%
. In Ref. \cite{dai2009number, dai2010approach}, the authors discuss the
mathematical basis of the boundary effects. Additionally, some related problems have also
been studied. In Ref. \cite{ozturk2009quantum}, the authors discuss the
thermal and potential conductivities by considering the size effects. In Ref.
\cite{su2010bose}, the authors considered the Bose--Einstein condensation of
the noninteracting bosons confined in a finite-size container. Based on the
size effect, an ideal heat engine has been constructed in the Ref.
\cite{nie2008performance, nie2009performance, nie2009quantum}.

In Sec. 2, based on the conformal mapping method and perturbation theory, we give the eigenvalue
of a point particle which moving freely and not being allowed to escape from the boundary in general 2-dimensional
regions. In Sec. 3, we provide the partition function of an ideal gas in a general 2-dimensional
regions based on the eigenvalue spectrum we calculate in the Sec. 2. In Sec. 4, we shall give some examples,
in which the numerical results and fitting results are given, and we compare the thermodynamic quantities of
the different regions. The Conclusion is given in Sec. 5.

\section{Conformal mapping method and perturbation theory}

In this section, we briefly introduce how to calculate the eigenvalue of a
point particle, that is moving freely and not being allowed to escape in
general 2-dimensional regions.

We consider a point particle freely moving in a general 2-dimensional region
$\mathcal{D}$ with complex boundary, namely a problem of 2-dimensional
infinite potential well. But the region $\mathcal{D}$ is a small deformation
of another 2-dimensional region $\Omega$ of which we have an exact solution. So we
can deal with this problem by way of the conformal mapping method and perturbation
theory in Ref. \cite{amore2010spectroscopy, segel1961application, segel1961application2}.

\subsection{Conformal mapping}

We can describe a point particle moving in a region $\mathcal{D}$ by using the
Schr\"{o}dinger equation with Dirichlet boundary condition,%

\begin{equation}
-\frac{\hbar^{2}}{2m}\left[  \frac{\partial^{2}}{\partial u^{2}}%
+\frac{\partial^{2}}{\partial v^{2}}\right]  \psi\left(  u,v\right)
=E\psi\left(  u,v\right)  ,
\end{equation}
For convenience, we set $\frac{\hbar^{2}}{2m}=1$. We transform the region
$\mathcal{D}$ onto region $\Omega$ by a holomorphic function, i.e. %

\begin{equation}
w=f\left(  z\right)  ,
\end{equation}
where $w=u+iv$, $z=x+iy$. According to the conformal mapping method we
mentioned in the introduction, after the conformal mapping $w=f\left(
z\right)  $, we obtain the equation of motion in the region $\Omega$, i.e.
\begin{equation}\label{traneq}
-\Sigma^{-1}\left[  \frac{\partial^{2}}{\partial x^{2}}+\frac{\partial^{2}%
}{\partial y^{2}}\right]  \psi\left(  x,y\right)  =E\psi\left(  x,y\right)  ,
\end{equation}
where $\Sigma=\left\vert \frac{df}{dz}\right\vert ^{2}$ is the conformal factor,
this equivalent local scaling factor after transformation.

The boundary condition can be simplified under the transformation, but at the
same time, the equation of motion is no longer represented by the Schr\"{o}dinger equation. The
Hamilton operator transforms as %

\begin{equation}
O=-\Sigma^{-1}\left[  \frac{\partial^{2}}{\partial x^{2}}+\frac{\partial^{2}%
}{\partial y^{2}}\right]  =-\Sigma^{-1}\triangle,
\end{equation}
where $\triangle=\frac{\partial^{2}}{\partial x^{2}}+\frac{\partial^{2}%
}{\partial y^{2}}$ is Laplace operator. In order to continue our calculation of
the spectrum, we require the use of perturbation theory.

\subsection{Perturbation theory}

In this section, based on the perturbation theory, we deal with the
transformed equation (\ref{traneq}).

The operator $O$ can be written as a symmetrized
form \cite{amore2010spectroscopy},
\begin{equation}
O=\frac{1}{\sqrt{\Sigma}}\left(  -\triangle\right)  \frac{1}{\sqrt{\Sigma}},
\label{symmetrizedform}%
\end{equation}
If the transformation we considered is a small deformation, then we can solve
the problem by way of perturbation theory.

For a small deformation, we rewrite the conformal factor as follows,
\begin{equation}
\Sigma\rightarrow\Sigma_{\eta}=1+\eta\sigma,\label{conformalfactor}%
\end{equation}
where $\sigma=\Sigma-1$, $\eta$ is a parameter denoting the order of the
perturbation. Substituting Eq. (\ref{conformalfactor}) into Eq. (\ref{symmetrizedform}%
), we expand the operator $O$ in powers of $\eta$, i.e.
\begin{equation}
O=O_{0}+\eta O_{1}+\eta^{2}O_{2}+....,
\end{equation}
where the coefficients of each power of $\eta$,
\begin{equation}
O_{0}=-\triangle,
\end{equation}%
\begin{equation}
O_{1}=-\frac{1}{2}\left[  \sigma\left(  -\triangle\right)  +\left(
-\triangle\right)  \sigma\right]  ,
\end{equation}%
\begin{equation}
O_{2}=\frac{1}{8}\left[  2\sigma\left(  -\triangle\right)  \sigma+3\sigma
^{2}\left(  -\triangle\right)  +3\left(  -\triangle\right)  \sigma^{2}\right]
,
\end{equation}%
\begin{equation}
O_{3}=-\frac{3}{16}\left[  \sigma^{2}\left(  -\triangle\right)  \sigma
+\sigma\left(  -\triangle\right)  \sigma^{2}\right]  -\frac{5}{16}\left[
\sigma^{3}\left(  -\triangle\right)  +\left(  -\triangle\right)  \sigma
^{3}\right]  ,
\end{equation}%
\[
\cdots.
\]
Using perturbation theory, we may calculate the eigenvalue of operator
$O$, the first two order read
\begin{equation}
E_{n}^{\left(  0\right)  }=\epsilon_{n}=\epsilon_{n_{x},n_{y}},
\end{equation}%
\begin{equation}
E_{n}^{\left(  1\right)  }=-\left\langle n\right\vert O_{1}\left\vert
n\right\rangle =-\epsilon_{n_{x},n_{y}}\left\langle n_{x},n_{y}\right\vert
\sigma\left\vert n_{x},n_{y}\right\rangle ,\label{firstorder}%
\end{equation}
where $\left\vert n\right\rangle =\left\vert n_{x},n_{y}\right\rangle $ is the
eigenfunction of zeroth order. We consider the correction to the eigenvalue up
to first order,
\begin{equation}
E_{n_{x},n_{y}}=\epsilon_{n_{x},n_{y}}\left(  1-\left\langle n_{x}%
,n_{y}\right\vert \sigma\left\vert n_{x},n_{y}\right\rangle \right)
.\label{eigenvalue}%
\end{equation}

Based on the conformal mapping method and perturbation theory in the
Ref. \cite{amore2010spectroscopy}, we obtain the perturbation expansion of the
spectrum of a point particle moving in a general 2-dimensional region. The
following section we will deal with the statistical problem of an ideal gas in a general
2-dimensional region based on the result attained in this section.

\section{Statistical problem of ideal gas}

We already obtained the spectrum of a point particle moving in a general
2-dimensional regions in the last section by way of the conformal mapping method and
perturbation theory. In this section, we consider the statistical problem of
the ideal gas.

Let us consider a system of ideal gas within a general 2-dimensional regions
$\mathcal{D}$. By definition we have the partition function \cite{pathria}, i.e. %
\begin{equation}
Z\left(  \beta\right)  =\sum_{\epsilon}e^{-\beta\epsilon}=\sum_{n_{x}}%
\sum_{n_{y}}e^{-\beta E_{n_{x},n_{y}}}.\label{partitionfunction}%
\end{equation}
Substituting Eq. (\ref{eigenvalue}) into Eq. (\ref{partitionfunction}), we obtain
\begin{equation}
Z\left(  \beta\right)  =\sum_{n_{x}}\sum_{n_{y}}\exp\left[  -\beta
\epsilon_{n_{x},n_{y}}\left(  1-\left\langle n_{x},n_{y}\right\vert
\sigma\left\vert n_{x},n_{y}\right\rangle \right)  \right]
.\label{partitionfunction2}%
\end{equation}
Furthermore, we can calculate several thermodynamic quantities using the partition function,
such as internal energy $U=-\frac{\partial}{\partial\beta}\ln Z\left(
\beta\right)  $ \cite{pathria}.

Nevertheless, it is difficult to calculate the sum in the expression of partition
function Eq. (\ref{partitionfunction2}). We can adopt of numerical method to
calculate the partition function in the specific case. The following, we will
give some examples.

\section{Example}

\begin{figure}[h]
  \centering
 \includegraphics[width=0.5\textwidth]{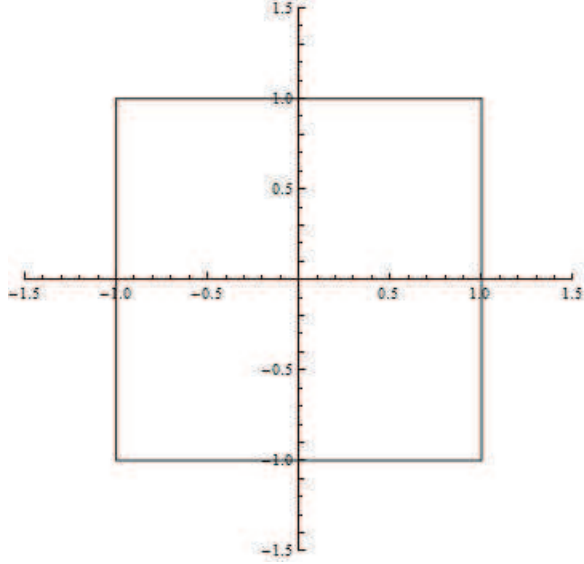}\\
  \caption{The square region}\label{fig0}
\end{figure}

In this section, we consider some specific examples. The examples considered in the following
are  an exact solution--- a solution with a square region, i.e. Fig. [\ref{fig0}],
The statistical problem of the 2-dimensional ideal gas in this region can be
solved precisely. We have the spectrum on the square region,
\begin{equation}
E_{n}=\epsilon_{n}=\epsilon_{n_{x},n_{y}}=\frac{\pi^{2}}{4}\left(  n_{x}%
^{2}+n_{y}^{2}\right)  ,\label{0eigenvalue}%
\end{equation}
and the corresponding wave function is
\begin{equation}
\left\vert n\right\rangle =\left\vert n_{x},n_{y}\right\rangle =\sin\left[
\frac{n_{x}\pi}{2}\left(  x+1\right)  \right]  \sin\left[  \frac{n_{y}\pi}%
{2}\left(  y+1\right)  \right]  .\label{0wavefunction}%
\end{equation}
We treat the result of the square region as the zeroth order when we deal with
the following examples. We can obtain the partition function of ideal gas on
this region, i.e.
\begin{align}
Z\left(  \beta\right)   &  =\sum_{n_{x}}\sum_{n_{y}}\exp\left[  -\beta
\frac{\pi^{2}}{4}\left(  n_{x}^{2}+n_{y}^{2}\right)  \right]  \nonumber\\
&  =\frac{1}{4}\left[  \vartheta_{3}\left(  0,e^{-\frac{1}{4}\pi^{2}\beta
}\right)  -1\right]  ^{2},
\end{align}
where $\vartheta_{3}\left(  0,z\right)  $ is Jacobi theta
function \cite{abramowitz2012handbook}.

We can also calculate several thermodynamic quantities such as the internal energy using the partition
function, i.e.\cite{pathria} %
\begin{align}
U  & =-\frac{\partial}{\partial\beta}\ln Z\left(  \beta\right)  \nonumber\\
& =\frac{\pi^{2}e^{-\frac{\pi^{2}\beta}{4}}\vartheta_{3}^{(0,1)}\left(
0,e^{-\frac{\pi^{2}\beta}{4}}\right)  }{2\left[  \vartheta_{3}\left(
0,e^{-\frac{\pi^{2}\beta}{4}}\right)  -1\right]  }.
\end{align}

The specific heat of the system is given by \cite{pathria}, i.e.
\begin{align}
C_{V} &  =\frac{\partial}{\partial T}U=\frac{\partial}{\partial\left(
\frac{1}{\beta}\right)  }U=\frac{\partial}{\left(  -\frac{1}{\beta^{2}%
}\right)  \partial\beta}U\nonumber\\
&  =\frac{e^{-\frac{\pi^{2}\beta}{2}}\pi^{4}\beta^{2}}{8\left[  \vartheta
_{3}\left(  0,e^{-\frac{\pi^{2}\beta}{4}}\right)  -1\right]  ^{2}}\left\{
-\vartheta_{3}^{(0,1)}\left(  0,e^{-\frac{\pi^{2}\beta}{4}}\right)
^{2}+e^{\frac{\pi^{2}\beta}{4}}\left[  \vartheta_{3}\left(  0,e^{-\frac
{\pi^{2}\beta}{4}}\right)  -1\right]  \vartheta_{3}^{(0,1)}\left(
0,e^{-\frac{\pi^{2}\beta}{4}}\right)  \right.  \nonumber\\
&  +\left.  \left[  \vartheta_{3}\left(  0,e^{-\frac{\pi^{2}\beta}{4}}\right)
-1\right]  \vartheta_{3}^{(0,2)}\left(  0,e^{-\frac{\pi^{2}\beta}{4}}\right)
\right\}  .
\end{align}

In the sections that follow we shall consider various cases.

\subsection{Transformation of $w=f\left(  z\right)  =z+\lambda z^{2}$}

\begin{figure}[h]
  \centering
  \includegraphics[width=0.5\textwidth]{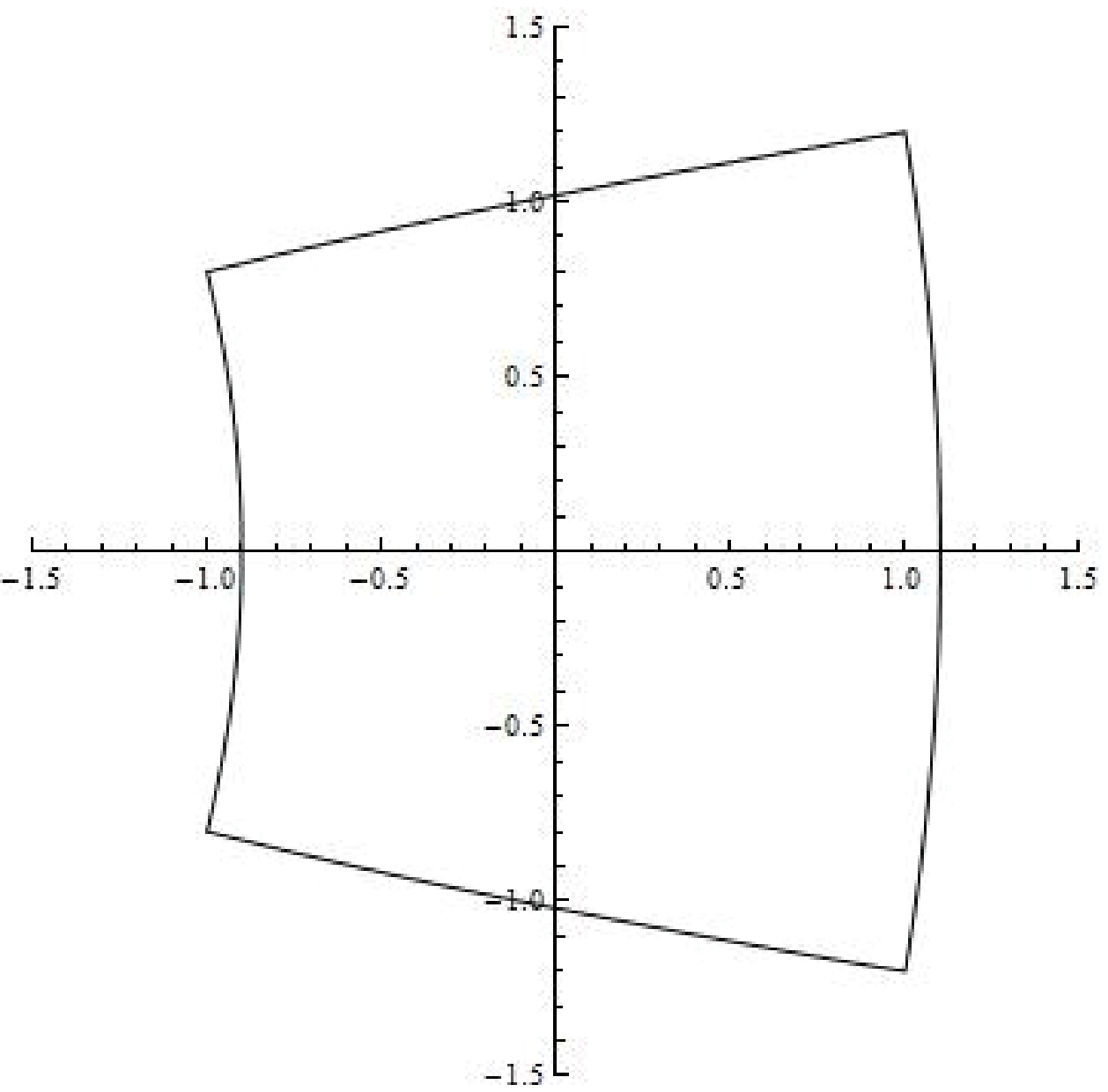}\\
  \caption{$\lambda=0.1$}\label{eg1}
\end{figure}

Let us consider the 2-dimensional regions $\mathcal{D}_{1}$ with complex
boundary, Fig. [\ref{eg1}]
\noindent where we set $\lambda=0.1$. After conformal transformation, we map the regions
$\mathcal{D}_{1}$ onto regions $\Omega$ Fig. [\ref{fig0}].

\subsubsection{Spectrum of eigenvalue}

First, we calculate the conformal factor of the transformation of $w=f\left(
z\right)  =z+\lambda z^{2}$, i.e.
\begin{equation}
\Sigma=\left\vert \frac{df}{dz}\right\vert ^{2}=1+4x\lambda+4\lambda
^{2}\left(  x^{2}+y^{2}\right)  ,
\end{equation}
then we have
\begin{align}
\sigma &  =\Sigma-1\nonumber\\
&  =4\lambda x+4\lambda^{2}\left(  x^{2}+y^{2}\right)
\end{align}
Substituting $\sigma$, zero order eigenvalue Eq. (\ref{0eigenvalue}) and zero
order eigenfunction Eq. (\ref{0wavefunction}) into the expression of first order
correction to eigenvalue Eq. (\ref{firstorder}), we obtain
\begin{align}
E_{n}^{\left(  1\right)  } &  =-\frac{\pi^{2}}{4}\left(  n_{x}^{2}+n_{y}%
^{2}\right)  \int_{-1}^{1}dx\int_{-1}^{1}dy\sin^{2}\left[  \frac{n_{x}\pi}%
{2}\left(  x+1\right)  \right]  \sin^{2}\left[  \frac{n_{y}\pi}{2}\left(
y+1\right)  \right]  \left[  4\lambda x+4\lambda^{2}\left(  x^{2}%
+y^{2}\right)  \right]  \nonumber\\
&  =-\frac{\pi^{2}}{4}\left(  n_{x}^{2}+n_{y}^{2}\right)  \frac{8}{3}%
\lambda^{2}\left(  1-\frac{3}{\pi^{2}n_{x}^{2}}-\frac{3}{\pi^{2}n_{y}^{2}%
}\right)  .
\end{align}

\subsubsection{Partition function}

Now, we calculate the partition function of 2-dimensional ideal gas on the
region $\mathcal{D}_{1}$. Substituting the eigenvalue into
Eq. (\ref{partitionfunction2}), we have
\begin{align}
Z\left(  \beta\right)   &  =\sum_{\epsilon}e^{-\beta\epsilon}\nonumber\\
&  =\sum_{n_{x}}\sum_{n_{y}}\exp\left\{  -\beta\frac{\pi^{2}}{4}\left(
n_{x}^{2}+n_{y}^{2}\right)  \left[  1-\frac{8}{3}\lambda^{2}\left(  1-\frac
{3}{\pi^{2}n_{x}^{2}}-\frac{3}{\pi^{2}n_{y}^{2}}\right)  \right]  \right\}
\end{align}
Obviously, it is hard to give an analytical result of the sum, so we can
calculate the numerical result of the sum for different $\beta$ (namely
different temperature). We calculate the partition function by choosing remove
different values of $\beta$. We list numerical results in the Tab. [\ref{tab1}],
\begin{center}
\begin{table}[h]
\caption{Numerical results}\label{tab1}
\centering
\begin{tabular}{|c|c|c|c|c|c|c|c|}
\hline
$\beta$ & 0.01 & 0.02 & 0.03 & 0.04 & 0.05 & 0.06 & 0.07  \\
\hline
$Z\left(\beta\right)$ & 27.158 & 12.5049 & 7.80701 & 5.5305 & 4.20164 & 3.33767 & 2.73481 \\
\hline

$\beta$  & 0.08 & 0.09 & 0.10 & 0.20 & 0.30 & 0.40 & 0.50  \\
\hline
$Z\left(\beta\right)$ & 2.29255 & 1.95574 & 1.69171 & 0.594704 & 0.288229 & 0.15799 & 0.0917356\\
\hline

$\beta$  & 0.60 & 0.70 & 0.80 & 0.90 & 1.00 & &\\
\hline
$Z\left(\beta\right)$ & 0.0547901 & 0.0331811 & 0.0202314 & 0.0123765& 0.00758345& & \\
\hline
\end{tabular}
\end{table}
\end{center}

In principle, we can give the numerical result for different temperature. But
we still give an expression of the partition function based on the table of
numerical results. The perturbation calculation is based on the square region
$\Omega$, so we can set the partition function on the region $\mathcal{D}_{1}$
as same as the form of the partition function on the square region $\Omega$,
\begin{equation}
Z\left(  \beta\right)  =\frac{m}{4}\left[  \vartheta_{3}\left(  0,e^{-\frac
{n}{4}\pi^{2}\beta}\right)  -1\right]  ^{2}.
\end{equation}
Taking the values from the Tab. [\ref{tab1}], we give the fitting value of ($m,n$), as
\[
m\rightarrow1.0317,\text{ \ \ \ \ \ \ }n\rightarrow1.004
\]
so we obtain the partition function of 2-dimensional ideal gas on the region
$\mathcal{D}_{1}$,
\begin{equation}
Z\left(  \beta\right)  =\frac{1.0317}{4}\left[  \vartheta_{3}\left(
0,e^{-\frac{1.004}{4}\pi^{2}\beta}\right)  -1\right]  ^{2}.
\end{equation}
The internal energy of this system is given by \cite{pathria}, i.e.%
\begin{align}
U  & =-\frac{\partial}{\partial\beta}\ln Z\left(  \beta\right)  \nonumber\\
& =\frac{1.004\pi^{2}e^{-\frac{1.004}{4}\pi^{2}\beta}\vartheta_{3}%
^{(0,1)}\left(  3,0,e^{-\frac{1.004}{4}\pi^{2}\beta}\right)  }{2\left[
\vartheta_{3}\left(  0,e^{-\frac{1.004}{4}\pi^{2}\beta}\right)  -1\right]  }.
\end{align}
We also have the specific heat of the system \cite{pathria}, i.e.
\begin{align}
C_{V} &  =\frac{\partial}{\partial T}U=\frac{\partial}{\partial\left(
\frac{1}{\beta}\right)  }U=\frac{\partial}{\left(  -\frac{1}{\beta^{2}%
}\right)  \partial\beta}U\nonumber\\
&  =\frac{1.004^{2}e^{-\frac{1.004}{2}\pi^{2}\beta}\pi^{4}\beta^{2}}{8\left[
\vartheta_{3}\left(  0,e^{-\frac{1.004}{4}\pi^{2}\beta}\right)  -1\right]
^{2}}\left\{  -\vartheta_{3}^{(0,1)}\left(  0,e^{-\frac{1.004}{4}\pi^{2}\beta
}\right)  ^{2}+\left[  \vartheta_{3}\left(  0,e^{-\frac{1.004}{4}\pi^{2}\beta
}\right)  -1\right]  \vartheta_{3}^{(0,2)}\left(  0,e^{-\frac{1.004}{4}\pi
^{2}\beta}\right)  \right.  \nonumber\\
&  \left.  +e^{\frac{1.004}{4}\pi^{2}\beta}\left[  \vartheta_{3}\left(
0,e^{-\frac{1.004}{4}\pi^{2}\beta}\right)  -1\right]  \vartheta_{3}%
^{(0,1)}\left(  0,e^{-\frac{1.004}{4}\pi^{2}\beta}\right)  \right\}  .
\end{align}

\begin{figure}[h]
  \centering
  \subfigure[The internal energy]{\label{U1}
                \includegraphics[width=0.45\textwidth]{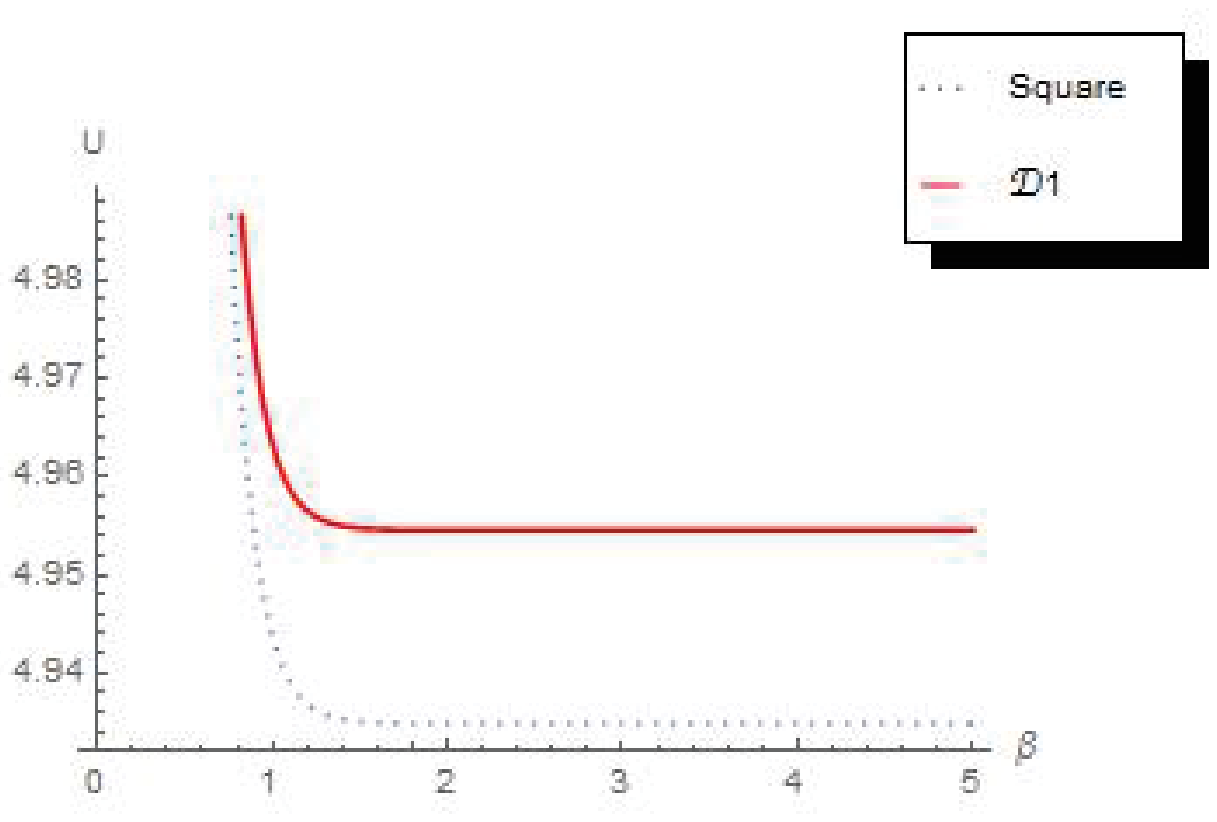}}
  \subfigure[The specific heat]{\label{C1}
                \includegraphics[width=0.45\textwidth]{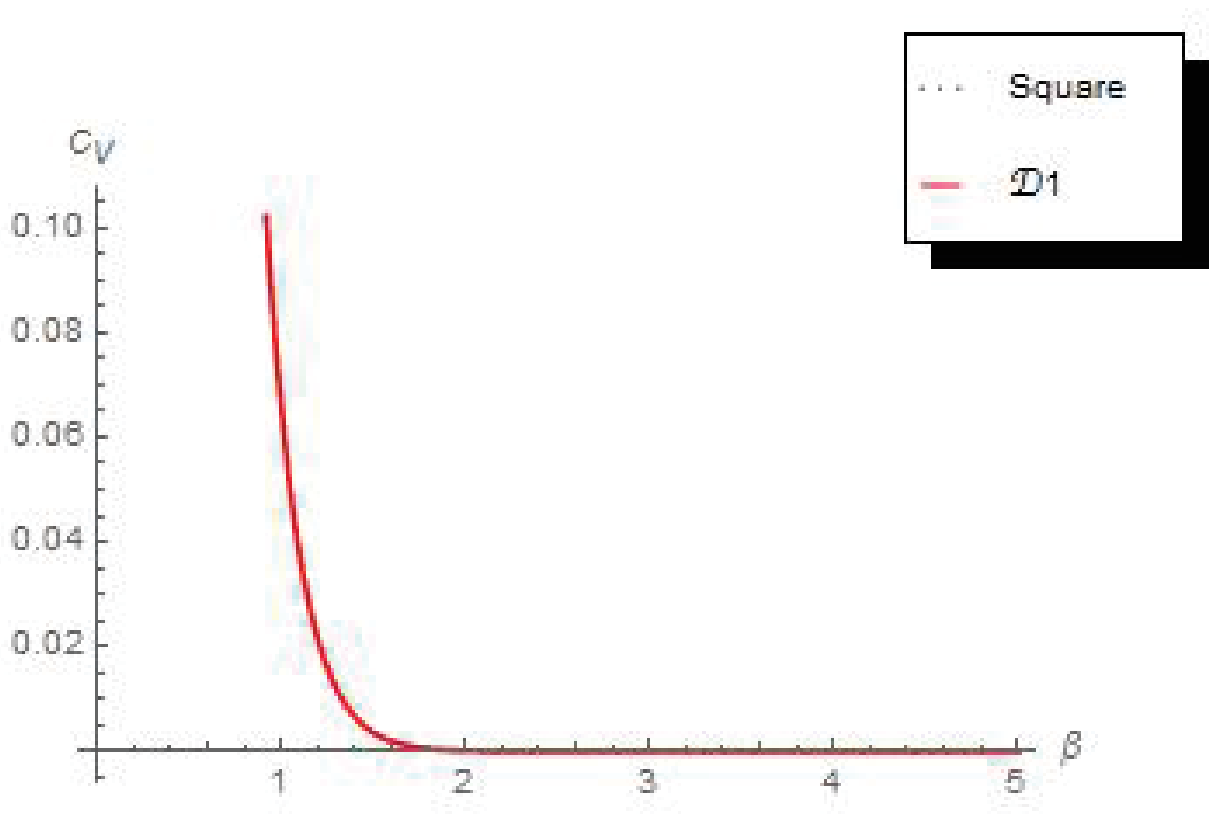}}
  \caption{The solid line represent the thermodynamic quantities of the square region, and the dotted line
 represent the thermodynamic quantities of the $\mathcal{D}_{1}$.}\label{UC1}
\vspace{\baselineskip}
\end{figure}

The effect of the thermodynamic quantities after the transformation are shown in the
Fig.[\ref{UC1}],

\subsection{Transformation of $w=f\left(  z\right)  =z+\lambda e^{z}$}
\begin{figure}[h]
  \centering
  \includegraphics[width=0.5\textwidth]{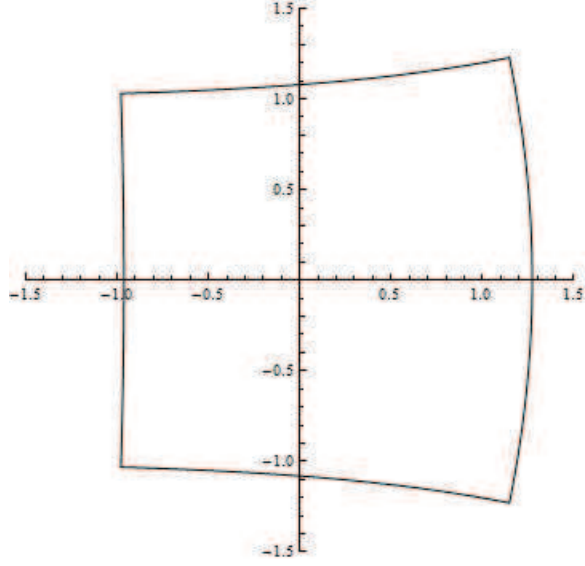}\\
  \caption{$\lambda=0.1$}\label{fig2}
\end{figure}

Let us consider the 2-dimensional regions $\mathcal{D}_{2}$ with complex
boundary Fig. [\ref{fig2}],
\noindent where we set $\lambda=0.1$. After conformal transformation, we map the regions
$\mathcal{D}_{2}$ onto regions $\Omega$ Fig. [\ref{fig0}].

\subsubsection{Spectrum of eigenvalue}

First, we calculate the conformal factor of transformation of $w=f\left(
z\right)  =z+\lambda e^{z}$, i.e.
\begin{equation}
\Sigma=\left\vert \frac{df}{dz}\right\vert ^{2}=1+2e^{x}\lambda\cos
y+\lambda^{2}e^{2x},
\end{equation}
then we have
\begin{align}
\sigma &  =\Sigma-1\nonumber\\
&  =2e^{x}\lambda\cos y+\lambda^{2}e^{2x}.
\end{align}
Substituting $\sigma$, zero order eigenvalue Eq. (\ref{0eigenvalue}) and zero
order eigenfunction Eq. (\ref{0wavefunction}) into the expression of first order
correction to eigenvalue Eq. (\ref{firstorder}), we obtain
\begin{align}
E_{n}^{\left(  1\right)  }  &  =-\frac{\pi^{2}}{4}\left(  n_{x}^{2}+n_{y}%
^{2}\right)  \int_{-1}^{1}dx\int_{-1}^{1}dy\sin^{2}\left[  \frac{n_{x}\pi}%
{2}\left(  x+1\right)  \right]  \sin^{2}\left[  \frac{n_{y}\pi}{2}\left(
y+1\right)  \right]  \left(  2\lambda e^{x}\cos y+\lambda^{2}e^{2x}\right)
\nonumber\\
&  =-\frac{\pi^{2}}{4}\left(  n_{x}^{2}+n_{y}^{2}\right)  \left[  \lambda
\frac{\left(  e-e^{-1}\right)  \pi^{4}n_{x}^{2}n_{y}^{2}\sin1}{\left(  \pi
^{2}n_{x}^{2}+1\right)  \left(  \pi^{2}n_{y}^{2}-1\right)  }+\lambda^{2}%
\frac{\pi^{2}n_{x}^{2}\sinh2}{2\left(  \pi^{2}n_{x}^{2}+4\right)  }\right]  .
\end{align}

\subsubsection{Partition function}

Now, we calculate the partition function of 2-dimensional ideal gas on the
region $\mathcal{D}_{2}$. Substituting the eigenvalue into
Eq. (\ref{partitionfunction2}), we have
\begin{align}
Z\left(  \beta\right)   &  =\sum_{\epsilon}e^{-\beta\epsilon}\nonumber\\
&  =\sum_{n_{x}}\sum_{n_{y}}\exp\left[  -\beta\frac{\pi^{2}}{4}\left(
n_{x}^{2}+n_{y}^{2}\right)  \left(  1-\lambda\frac{\left(  e-e^{-1}\right)
\pi^{4}n_{x}^{2}n_{y}^{2}\sin1}{\left(  \pi^{2}n_{x}^{2}+1\right)  \left(
\pi^{2}n_{y}^{2}-1\right)  }-\lambda^{2}\frac{\pi^{2}n_{x}^{2}\sinh2}{2\left(
\pi^{2}n_{x}^{2}+4\right)  }\right)  \right]  .
\end{align}
Sililar to case 1, we calculate the numerical result of the sum for different
$\beta$ and give the numerical results in the Tab. [\ref{tab2}],
\begin{center}
\begin{table}[h]
\caption{Numerical results}\label{tab2}
\centering
\begin{tabular}{|c|c|c|c|c|c|c|c|}
\hline
$\beta$ & 0.01 & 0.02 & 0.03 & 0.04 & 0.05 & 0.06 & 0.07  \\
\hline
$Z\left(\beta\right)$ & 34.458 & 16.0312 & 10.0942 &7.20541 & 5.51287 & 4.40864 & 3.63556 \\
\hline

$\beta$  & 0.08 & 0.09 & 0.10 & 0.20 & 0.30 & 0.40 & 0.50  \\
\hline
$Z\left(\beta\right)$ & 3.06661& 2.63197 &2.29021 & 0.852253 & 0.437876 & 0.255918 & 0.15969\\
\hline

$\beta$  & 0.60 & 0.70 & 0.80 & 0.90 & 1.00 & &\\
\hline
$Z\left(\beta\right)$ & 0.103308 & 0.0682117 & 0.0455631 & 0.0306348 & 0.020674 & & \\
\hline

\end{tabular}
\end{table}
\end{center}
Taking the values from the Tab. [\ref{tab2}], we obtain the partition function of
2-dimensional ideal gas on the region $\mathcal{D}_{2}$,
\begin{equation}
Z\left(  \beta\right)  =\frac{1.00614}{4}\left[  \vartheta_{3}\left(
0,e^{-\frac{0.788871}{4}\pi^{2}\beta}\right)  -1\right]  ^{2}.
\end{equation}
The internal energy of this system is given by \cite{pathria}, i.e. %
\begin{align}
U  & =-\frac{\partial}{\partial\beta}\ln Z\left(  \beta\right)  \nonumber\\
& =\frac{0.788871\pi^{2}e^{-\frac{0.788871}{4}\pi^{2}\beta}\vartheta
_{3}^{(0,1)}\left(  3,0,e^{-\frac{0.788871}{4}\pi^{2}\beta}\right)  }{2\left[
\vartheta_{3}\left(  0,e^{-\frac{0.788871}{4}\pi^{2}\beta}\right)  -1\right]
}.
\end{align}

\begin{figure}[h]
  \centering
  \subfigure[The internal energy]{\label{U2}
                \includegraphics[width=0.45\textwidth]{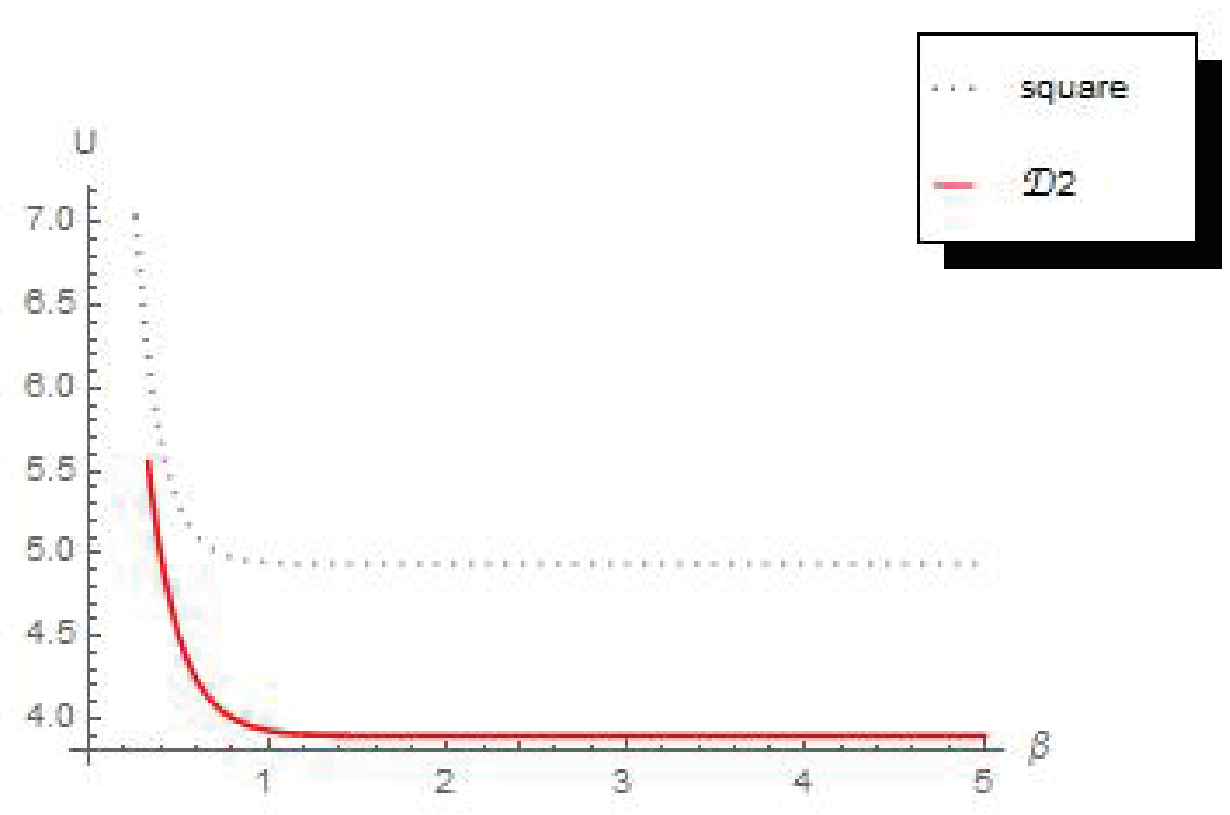}}
  \subfigure[The specific heat]{\label{C2}
                \includegraphics[width=0.45\textwidth]{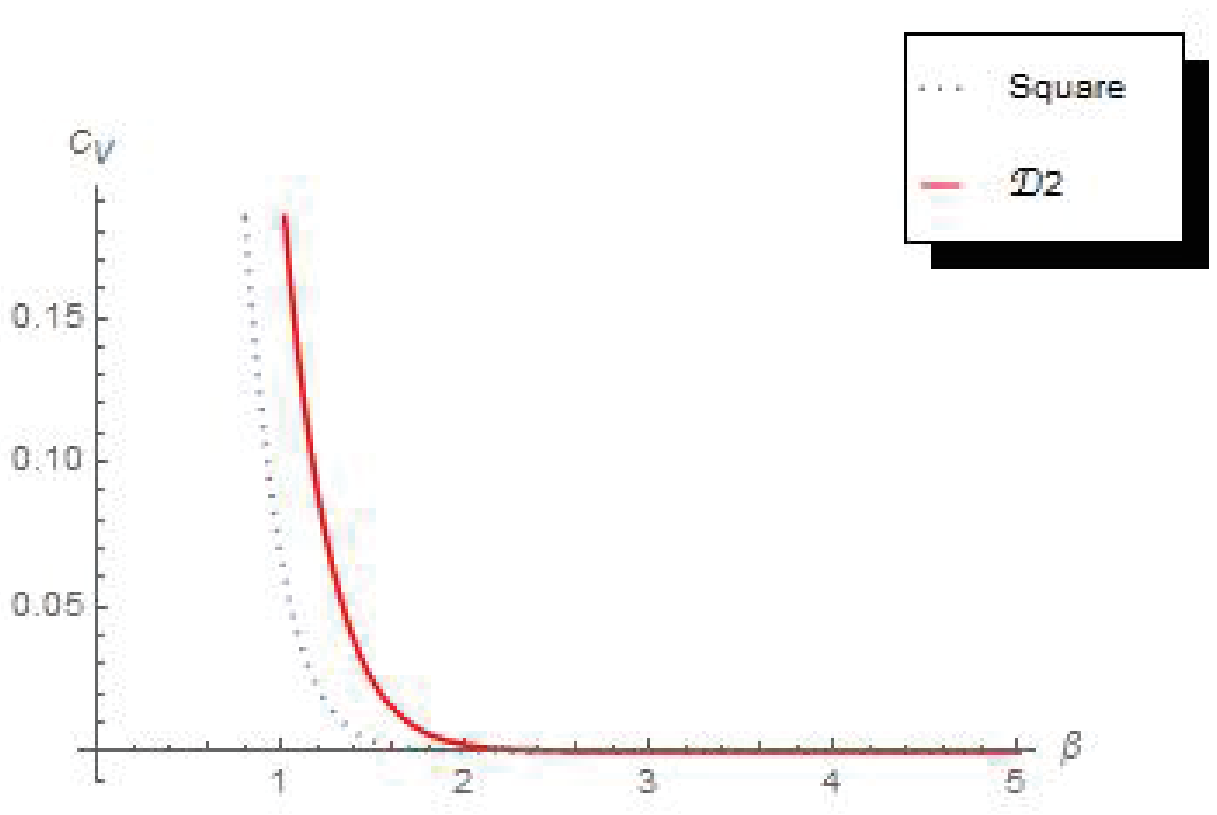}}
  \caption{The solid line represent the thermodynamic quantities of the square region, and the dotted line
 represent the thermodynamic quantities of the $\mathcal{D}_{2}$.}\label{UC2}
\vspace{\baselineskip}
\end{figure}

We also have the specific heat of the system \cite{pathria}, i.e.
\begin{align}
C_{V} &  =\frac{\partial}{\partial T}U=\frac{\partial}{\partial\left(
\frac{1}{\beta}\right)  }U=\frac{\partial}{\left(  -\frac{1}{\beta^{2}%
}\right)  \partial\beta}U\nonumber\\
&  =\!\frac{0.788871^{2}\!e^{-\frac{0.788871}{2}\pi^{2}\beta}\pi^{4}\beta^{2}%
}{8\!\left[  \!\vartheta_{3}\left(  0,e^{-\frac{0.788871}{4}\pi^{2}\beta
}\right)  \!-1\!\right]  \!^{2}}\left\{  \!-\!\vartheta_{3}^{(0,1)}\left(
0,\!e^{-\frac{0.788871}{4}\pi^{2}\beta}\right)  ^{2}\!+\!\left[
\!\vartheta_{3}\left(  0,e^{-\frac{0.788871}{4}\pi^{2}\beta}\right)
\!-1\!\right]  \!\vartheta_{3}^{(0,2)}\!\left(  \!0,\!e^{-\frac{0.788871}%
{4}\pi^{2}\beta}\!\right)  \!\right.  \nonumber\\
&  \left.  +e^{\frac{0.788871}{4}\pi^{2}\beta}\left[  \vartheta_{3}\left(
0,e^{-\frac{0.788871}{4}\pi^{2}\beta}\right)  -1\right]  \vartheta_{3}%
^{(0,1)}\left(  0,e^{-\frac{0.788871}{4}\pi^{2}\beta}\right)  \right\}  .
\end{align}
The effect of the thermodynamic quantities after the transformation are shown in the
Fig.[\ref{UC2}],

\subsection{Transformation of $w=f\left(  z\right)  =z+\lambda\sin z$}
\begin{figure}[h]
  \centering
  \includegraphics[width=0.5\textwidth]{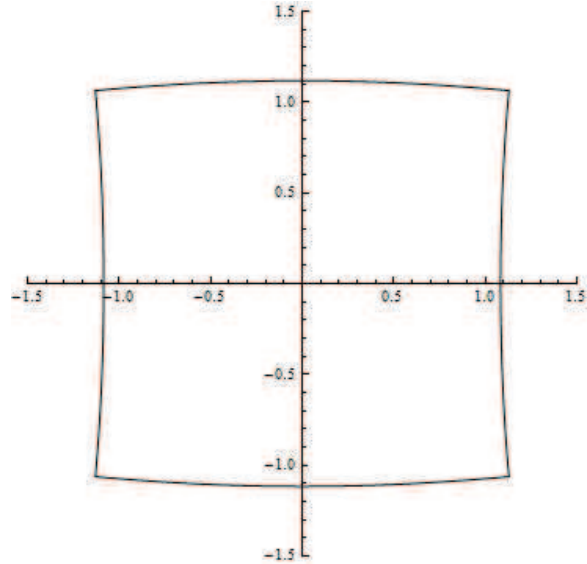}\\
  \caption{$\lambda=0.1$}\label{fig3}
\end{figure}

Finally, we consider the 2-dimensional regions $\mathcal{D}_{3}$ with complex
boundary Fig. [\ref{fig3}] ,
\noindent where we set $\lambda=0.1$. After conformal transformation, we map the regions
$\mathcal{D}_{3}$ onto regions $\Omega$ Fig. [\ref{fig0}].

\subsubsection{Spectrum of eigenvalue}

First, we calculate the conformal factor of the transformation of $w=f\left(
z\right)  =z+\lambda\sin z$, i.e.
\begin{equation}
\Sigma=\left\vert \frac{df}{dz}\right\vert ^{2}=1+2\lambda\cos x\cosh
y+\frac{1}{2}\lambda^{2}\left(  \cos2x+\cosh2y\right)  ,
\end{equation}
then we have
\begin{align}
\sigma &  =\Sigma-1\nonumber\\
&  =2\lambda\cos x\cosh y+\frac{1}{2}\lambda^{2}\left(  \cos2x+\cosh2y\right)
.
\end{align}
Substituting $\sigma$, zero order eigenvalue Eq. (\ref{0eigenvalue}) and zero
order eigenfunction Eq. (\ref{0wavefunction}) into the expression of first order
correction to eigenvalue Eq. (\ref{firstorder}), we obtain
\begin{align}
E_{n}^{\left(  1\right)  } &  =-\frac{\pi^{2}}{4}\left(  n_{x}^{2}+n_{y}%
^{2}\right)  \int_{-1}^{1}dx\int_{-1}^{1}dy\sin^{2}\left[  \frac{n_{x}\pi}%
{2}\left(  x+1\right)  \right]  \sin^{2}\left[  \frac{n_{y}\pi}{2}\left(
y+1\right)  \right] \nonumber\\
&  \times\left[  2\lambda\cos x\cosh y+\frac{1}{2}\lambda
^{2}\left(  \cos2x+\cosh2y\right)  \right]  \nonumber\\
&  =-\frac{\pi^{2}}{4}\left(  n_{x}^{2}+n_{y}^{2}\right)  \left[  \lambda
\frac{\left(  e-e^{-1}\right)  \pi^{4}n_{x}^{2}n_{y}^{2}\sin1}{\left(  \pi
^{2}n_{x}^{2}-1\right)  \left(  \pi^{2}n_{y}^{2}+1\right)  }+\lambda^{2}%
\frac{\pi^{2}n_{x}^{2}\sin2}{4\pi^{2}n_{x}^{2}-16}+\lambda^{2}\frac{\pi
^{2}n_{y}^{2}\sinh2}{4\pi^{2}n_{y}^{2}+16}\right]  .
\end{align}

\subsubsection{Partition function}

Now, we calculate the partition function of 2-dimensional ideal gas on the
region $\mathcal{D}_{3}$. Substituting the eigenvalue into
Eq. (\ref{partitionfunction2}), we have
\begin{align}
Z\left(  \beta\right)   &  =\sum_{\epsilon}e^{-\beta\epsilon}\nonumber\\
&  =\sum_{n_{x}}\sum_{n_{y}}\exp\left[  -\beta\frac{\pi^{2}}{4}\left(
n_{x}^{2}+n_{y}^{2}\right)  \left(  1-\lambda\frac{\left(  e-e^{-1}\right)
\pi^{4}n_{x}^{2}n_{y}^{2}\sin1}{\left(  \pi^{2}n_{x}^{2}-1\right)  \left(
\pi^{2}n_{y}^{2}+1\right)  }-\lambda^{2}\frac{\pi^{2}n_{x}^{2}\sin2}{4\pi
^{2}n_{x}^{2}-16}-\lambda^{2}\frac{\pi^{2}n_{y}^{2}\sinh2}{4\pi^{2}n_{y}%
^{2}+16}\right)  \right]
\end{align}
Same as case 1, we calculate the numerical result of the sum for different
$\beta$ and give the numerical results in the Tab. [\ref{tab3}],
\begin{center}
\begin{table}[h]
\caption{Numerical results}\label{tab3}
\centering
\begin{tabular}{|c|c|c|c|c|c|c|c|}
\hline
$\beta$ & 0.01 & 0.02 & 0.03 & 0.04 & 0.05 & 0.06 & 0.07  \\
\hline
$Z\left(\beta\right)$ & 34.1644 & 15.8955 & 10.0092 &7.14507 & 5.46694 & 4.37208 & 3.60555 \\
\hline

$\beta$  & 0.08 & 0.09 & 0.10 & 0.20 & 0.30 & 0.40 & 0.50  \\
\hline
$Z\left(\beta\right)$ & 3.0414 & 2.61042 & 2.27152 & 0.845474 &0.434401 & 0.25384 & 0.158328\\
\hline

$\beta$  & 0.60 & 0.70 & 0.80 & 0.90 & 1.00 & &\\
\hline
$Z\left(\beta\right)$ & 0.102364 & 0.0675327 & 0.0450658 & 0.0302677& 0.0204026& & \\
\hline
\end{tabular}
\end{table}
\end{center}
Taking the values from the Tab. [\ref{tab3}], we obtain the partition function of
2-dimensional ideal gas on the region $\mathcal{D}_{3}$,
\begin{equation}
Z\left(  \beta\right)  =\frac{0.996072}{4}\left[  \vartheta_{3}\left(
0,e^{-\frac{0.787783}{4}\pi^{2}\beta}\right)  -1\right]  ^{2}.
\end{equation}
The internal energy of this system is given by \cite{pathria}, i.e. %
\begin{align}
U  & =-\frac{\partial}{\partial\beta}\ln Z\left(  \beta\right)  \nonumber\\
& =\frac{0.787783\pi^{2}e^{-\frac{0.787783}{4}\pi^{2}\beta}\vartheta
_{3}^{(0,1)}\left(  3,0,e^{-\frac{0.787783}{4}\pi^{2}\beta}\right)  }{2\left[
\vartheta_{3}\left(  0,e^{-\frac{0.787783}{4}\pi^{2}\beta}\right)  -1\right]
}.
\end{align}

We also have the specific heat of the system \cite{pathria}, i.e.
\begin{align}
C_{V} &  =\frac{\partial}{\partial T}U=\frac{\partial}{\partial\left(
\frac{1}{\beta}\right)  }U=\frac{\partial}{\left(  -\frac{1}{\beta^{2}%
}\right)  \partial\beta}U\nonumber\\
&  \!=\!\frac{0.787783^{2}\!e^{-\frac{0.787783}{2}\pi^{2}\beta}\pi^{4}%
\beta^{2}}{8\!\left[  \!\vartheta_{3}\left(  0,\!e^{-\frac{0.787783}{4}\pi
^{2}\beta}\right)  -1\!\right]  ^{2}}\left\{  \!-\vartheta_{3}^{(0,1)}\left(
0,\!e^{-\frac{0.787783}{4}\pi^{2}\beta}\right)  ^{2}+\!\left[  \!\vartheta
_{3}\left(  0,\!e^{-\frac{0.787783}{4}\pi^{2}\beta}\!\right)  \!-1\!\right]
\!\vartheta_{3}^{(0,2)}\left(  0,\!e^{-\frac{0.787783}{4}\pi^{2}\beta}\right)
\right.  \nonumber\\
&  \left.  +e^{\frac{0.787783}{4}\pi^{2}\beta}\left[  \vartheta_{3}\left(
0,e^{-\frac{0.787783}{4}\pi^{2}\beta}\right)  -1\right]  \vartheta_{3}%
^{(0,1)}\left(  0,e^{-\frac{0.787783}{4}\pi^{2}\beta}\right)  \right\}  .
\end{align}

\begin{figure}[h]
  \centering
  \subfigure[The internal energy]{\label{U3}
                \includegraphics[width=0.45\textwidth]{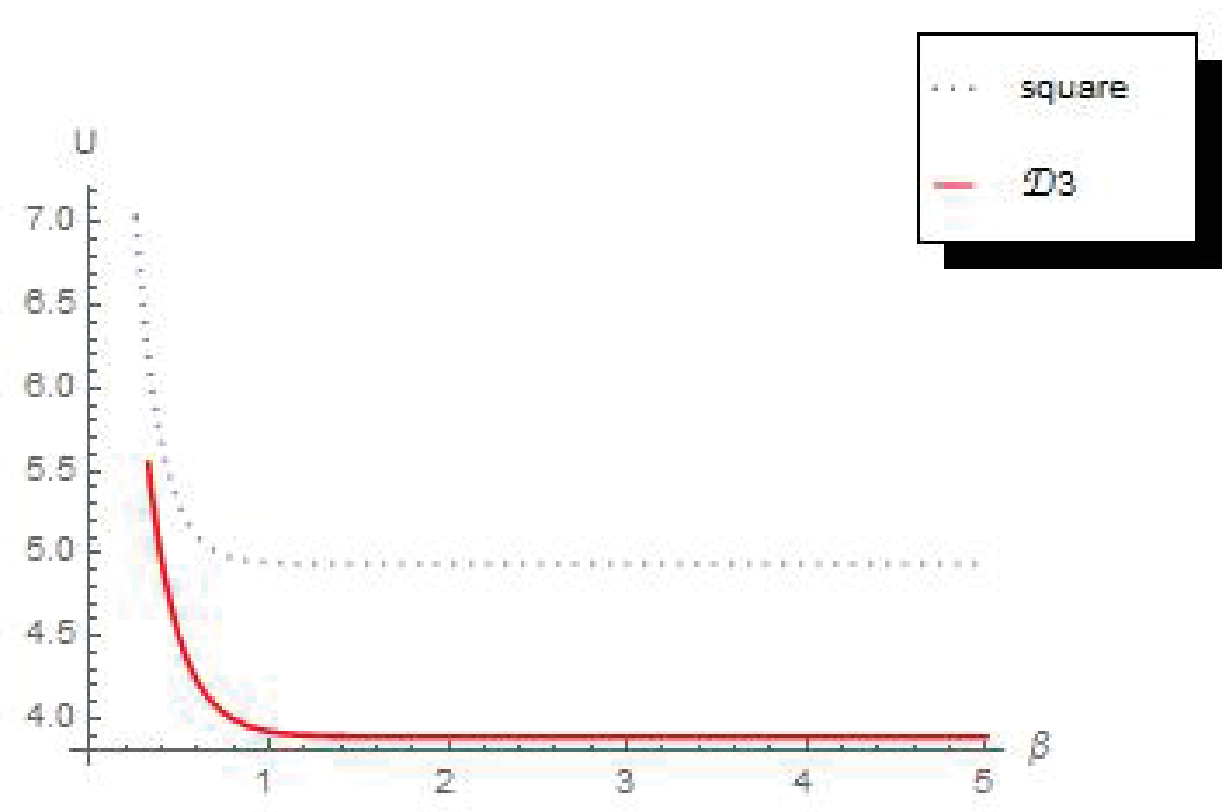}}
  \subfigure[The specific heat]{\label{C3}
                \includegraphics[width=0.45\textwidth]{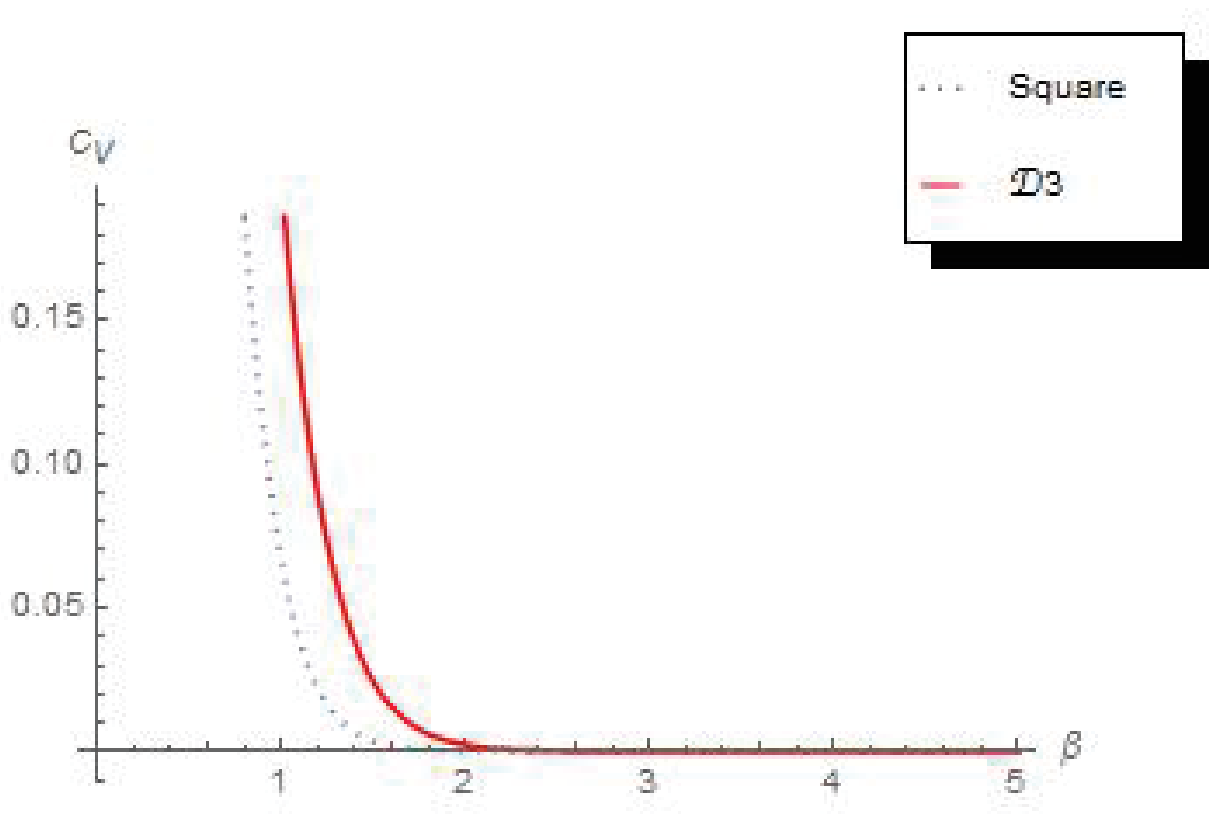}}
  \caption{The solid line represent the thermodynamic quantities of the square region, and the dotted line
 represent the thermodynamic quantities of the $\mathcal{D}_{3}$.}\label{UC3}
\vspace{\baselineskip}
\end{figure}

The effect of the thermodynamic quantities after the transformation are shown in the
Fig.[\ref{UC3}],
\section{Conclusion}

In this paper, based on the conformal mapping method and perturbation theory,
we present an approach to solve the statistical problem of ideal gas within
general 2-dimensional regions. We considered some example and give numerical
results. For more general case, we can state the transformation as follow:
\begin{equation}
f\left(  z\right)  =z+\sum\lambda_{n}z^{n}+\sum\alpha_{n}e^{nz}+\sum\beta
_{n}\sin nz.
\end{equation}
Then we obtain the transformation by calculating the undetermined parameters
$\lambda_{n}$, $\alpha_{n}$ and $\beta_{n}$. Using the approach given in this paper,
we are able to solve the statistical problem of ideal gas in general 2-dimensional regions.

\acknowledgments
This work is supported in part by Capital University of Economics and Business Specially commissioned projects.

% The bibliography will probably be heavily edited during typesetting.
% We'll parse it and, using the arxiv number or the journal data, will
% query inspire, trying to verify the data (this will probalby spot
% eventual typos) and retrive the document DOI and eventual errata.
% We however suggest to always provide author, title and journal data:
% in short all the informations that clearly identify a document.

\end{document}